\documentstyle[12pt,psfig]{l-aa}
\begin{document}
\def\msun{M_{\odot}}
\def\ergsec{\hbox{erg s$^{-1}$}}
\def\ergcmsec{\hbox{erg cm$^{-2}$ s$^{-1}$}}
\def\countsec{\hbox{counts s$^{-1}$}}
\def\photcmsec{\hbox{photon cm${-2}$ s$^{-1}$}}
\def\degmark{^\circ}
\def \rsun {\ifmmode$R$_{\odot}\else R$_{\odot}$\fi}
\thesaurus{08.02.1,08.05.3,08.14.1,08.16.7,13.25.1}
\title{An evolutionary model for  SAX J1808.4 -- 3658}
\author{Ene Ergma\inst{1,2} and Jelena Antipova\inst{1}}
\institute{Tartu University, Physics Department, \"Ulikooli 18,
  EE2400 Tartu, Estonia
\and
Astronomical Institute ``Anton Pannekoek'',Kruislaan 403,
 1098 SJ Amsterdam, The Netherlands}
\offprints{Ergma}
\date{Received;accepted}
\maketitle
\begin{abstract}
 An evolutionary scenario to explain the transient nature
 and short total duration of the X-ray burst of
 SAX J1808.4 -- 3658 
 is proposed. An optical companion  of the neutron
 star(a``turn-off'' Main - Sequence star)   fills
 its Roche lobe at the orbital period ($P_\textrm{orb}$) $\sim$ 19 hours.
During the initial high mass- transfer phase when   the neutron
 star is a persistent X-ray source, the neutron star is spun up to
 a millisecond period. Due to its chemical composition gradient,
 the secondary does not become fully convective when its mass
 decreases below 0.3 $\msun$, hence a magnetic braking remains an
 effective mechanism to remove orbital angular momentum and the system
  evolves with Roche - lobe overflow towards a short orbital period.
  Near an orbital period of two hours  the mass transfer rate becomes
 so small ($\sim$ $10^{-11}\msun$/yr) that the system can not continue
 to be observed as a persistent X-ray source. During further
 Roche - lobe filling evolution   deep mixing allows
the surface of secondary to become more and more  helium rich
 . Since the accreted matter is helium rich, it is easy to explain
 observed short total duration of the burst .  This evolutionary picture suggest
 that radio emission can be observed only at shorter wavelength's .
 Our model predicts a faster orbital period decay than expected if
 the orbital evolution is  driven only by  gravitational wave
 radiation.  
\end{abstract}
\section{Introduction}
The transient X-ray burster SAX J1808.4 -- 3658 was discovered
 in September 1996 with the BeppoSAX Wide Field Cameras 
 (in't Zand et al., 1998). Two very bright type I X-ray bursts with
  short decay times
  $\sim$ 8  s (2 -- 8 keV) separated by 14.61 hours were observed.
The double-peaked time history of both bursts at high energies
 suggests a peak luminosity close to the Eddington limit which
 implies a distance to this object of 4 kpc (in't Zand et al. 1998). 
The observed peak flux of the steady emission was about
 2$\times 10^{-9}$erg s$^{-1}$ cm$^{-2}$ at 2 -- 10 keV. For
 a power-law spectrum and a distance,  d = 4 kpc the luminosity
 is $\sim$ 6$\times 10^{36}$ erg s$^{-1}$ and for a bremsstrahlung
 spectrum the luminosity is $\sim$5$\times 10^{36}$erg s$^{-1}$.
 From these values the mean accretion rate is $\sim$ 3$\times 10^{16}$g
 s$^{-1}$ (assuming that the neutron star radius is equal to 10 km and
 the  mass is 1.4 $\msun$). Recently  a transient source
 XTE J1808 -- 365, positionally coincident with SAX J1808.4 -- 3658 to
 within a few arcmin, was detected with the Rossi X-ray  Timing Explorer
 . The 2 -- 10 keV X-ray flux ($F_\textrm{x}$) on 1998 April 11
 was 1.5$\times 10^{-9}$erg s$^{-1}$cm$^{-2}$( Marshall, 1998).
 Wijnands \& van der Klis (1998)  discovered a coherent 2.49 millisecond
 signal from this source and they  estimated an upper limit on the
 magnetic field strength B  of $\sim (2-6)$$\times 10^8$ Gauss.
 Gilfanov et al. (1998)  found that between April 25 and 29 1998
 this X-ray pulsar  showed an abrupt decline of the X-ray luminosity
 which they interpreted as a result of centrifugal inhibitation
  at the luminosity level of a few $\times 10^{35}$erg s$^{-1}$.
 From this they estimated an upper limit on the magnetic field strength
 of B$\leq$ 3.5$\times 10^7$ d(kpc) $(M/1.4\msun)^{1/3}$($F_\textrm{x}/10^{-9}
erg s^{-1}$$cm^{-2}$ )$
^{1/2}$$R_\textrm{6}^{-8/3}$ Gauss, which for d = 4 kpc
 and a  canonical neutron star mass (1.4 $\msun$) and radius (R= 10 km)
is 1.4 $\times 10^8$ Gauss.
Gilfanov et al. (1998) have estimated that
 the total energy of the 1998 outburst was $\approx$ 7.8$\times
 10^{42}$ erg (3 -- 150 keV), corresponding to an accreted mass
 of $\sim$$10^{-11}\msun$. Since the time between two ourburst was
 $\sim$ 1.5 years the average accretion rate is
 $\sim$$10^{-11}\msun$$yr^{-1}$.

 Chakrabarty \& Morgan (1998) found the source is a binary with
 orbital period of 2.01 hours, and a very small mass function
 f($\msun$)=3.85$\times 10^{-5}$$\msun$. Assuming a random distribution
 of binary inclinations, the a priori probability of observing a
 system with inclination $\it i$ or smaller is ( 1 -- cos ${\it i}$).
 For an assumed 1.35 $\msun$ pulsar , the 95\%-confidence upper limit
 on the secondary mass is $M_2$$<$0.14$\msun$ and for a 2 $\msun$ pulsar
 this limit is 0.18$\msun$. Chakrabarty \& Morgan(1998) discussed a
 possible evolutionary scenario of this system, and proposed that it
 is closely related to the ``black-widow'' millisecond radio pulsar(
 PSR 1957+20, PSR J2051 -- 0827) which are evaporating their companion
 through irradiation.

Filippenko \& Leonard (1998) have found that Keck spectra
 (range 554-685 nm) of the suspected optical counterpart
 (Roche et al. 1998, Giles et al. 1998) of SAX J1808.4-3658 reveal
 a possible very weak, double-peaked $H_{\rm \alpha}$ emission line.

In this short note we propose a model of the  evolutionary history
 for this system
which is able to explain the main  observed properties.

 \section{\bf Evolutionary picture}

Orbital period determination for low-mass X-ray binaries (LMXB)
have been a rather difficult task. Up to now $\sim$ 20 LMXB with the
 orbital periods less than one day are known ( van Paradijs,1995).
If we compare the distribution of the orbital periods of LMXB
and the systems with white dwarfs as accretor (catalysmic variable - CV)
then we can see that the orbital period distribution for two classes
(LMXB and CV) is different. For both classes the orbital
periods show a clustering between 3 -- 10 hours. There are no LMXB
in the CV period gap of 2 -- 3 hours but if the orbital period
distribution for CV shows second peak between 1 -- 2 hours there
 was no LMXB  between orbital periods from 50 min to 3 hours (White\&
Mason, 1985). So SAX J 1808.4--3658 is first transient X-ray
source discovered in this orbital period range.

Based on data for soft X-ray transients and persistent low-mass
 X-ray binaries with known distance and orbital period van Paradijs
 (1996) showed that for these systems the distributions in a
 ($P_\textrm{orb}$, $\dot{M}$) diagram can be understood with the dwarf
 nova instability criterion adapted to account for X-ray heating
 of the accretion disk. Soft X-ray  outbursts can occur if the mass
 transfer rate, $\dot{M}$, is below a critical value
 $\dot{M}^{irr}_\textrm{crit}$ (which depends on orbital period). 
 For neutron star low-mass X-ray binaries King et al (1996)
 have obtained the following formula for the critical mass accretion
 rate $\dot{M}^{irr}_\textrm{crit}$
\begin{equation}
\dot{M}_\textrm{crit}^{irr}\approx 5.0\times 10^{-11} M_\textrm{ns}^{2/3}
(\frac{P_\textrm{orb}}{3hrs})^{4/3} \msun yr^{-1}
\end{equation}
where $M_\textrm{ns}$ is the neutron star mass in units of solar mass.

The evolutionary history of an LMXB and CV depends on two main time scales: (i)
The hydrogen burning time scale $t_\textrm{nuc}$ and (ii) the angular momentum loss
 time scale $t_\textrm{am}$ by magnetic braking, gravitational radiation and by mass
 loss from the system.

{\bf First case}

 If the Roche lobe filling secondary is an unevolved main-sequence star 
then $t_\textrm{nuc}\gg t_\textrm{am}$. Near the orbital period
$\sim$ 3 hours the magnetic braking switches off and the secondary loses
 its Roche lobe contact.  Second Roche lobe filling takes place near the
 orbital period of 2 hours and mass exchange rate is now $\approx$ $10^ {-10}$
$M\sun$/yr. Since this mass accretion rate is larger than
 $\dot{M}^{irr}_\textrm{crit}$ the persistent X-ray sources with short 
orbital periods must be observed. 

In the period gap the evolutionary paths for CV and LMXB may greatly
differ. A neutron star with sufficiently weak magnetic fields accreting
for a long time from a surrounding Keplerian disk can be spun up to
 millisecond periods.
Van den Heuvel\&van Paradijs (1988) proposed that the evaporation of companion
 stars (induced by MSP) could account the lack of persistent low-mass
X-ray sources below the period gap.

{\bf Second case}

If  the Roche lobe filling secondary is an "turn--off" Main--Sequence star
 then$t_\textrm{am}$$ \sim$$t_\textrm{nuc}$. For this case  according to
 the results of
 computations (Tutukov et al 1985, Pylyser \& Savonije , 1989, Ergma et al ,
 1998 )
at first the mass exchange proceeds on the thermal time scale of
the secondary, but later $\dot{M}$ quickly decreases.  Due to 
a chemical composition gradient
 the secondary did not become fully convective
 when its mass had  decreased below 0.3 $\msun$. Magnetic braking is
 still the most effective mechanism to remove orbital angular momentum
 and the system evolves towards a short orbital period via Roche - lobe
overflow. In the period gap the mass accretion rate is very low ($\ll$ 
$10^ {-10}$$\msun$/yr) which is less than the critical mass accretion rate
for system to be persistent X-ray source.

First case allows easily to understand the evolution of the LMXB
with the orbital period between 3 and 10--12 hours.

Since SAX J1808.4 --3658 is transient X-ray source 
we propose that the progenitor SAX J1808.4 -- 3658 evolves
 according to second case . Using the
 evolutionary program of Sarna \& De Greve(1994,1996) we have calculated
 an evolutionary sequence for a binary with a 1.4$\msun$ neutron star
 and a 1$\msun$ companion.   The  secondary fills its Roche lobe a
 $P_\textrm{orb}$ = 0.8 days. When the orbital period is two hours  the
 $\dot{M}$, is 4.7$\times 10^{-11}$   $\msun$$yr^{-1}$ and the surface
 hydrogen abundance has been decreased to X = 0.3. As we will see later
this value is important to explain the short duration of the X-ray
 bursts. 

\begin{figure}
\begin{center}
\psfig{figure=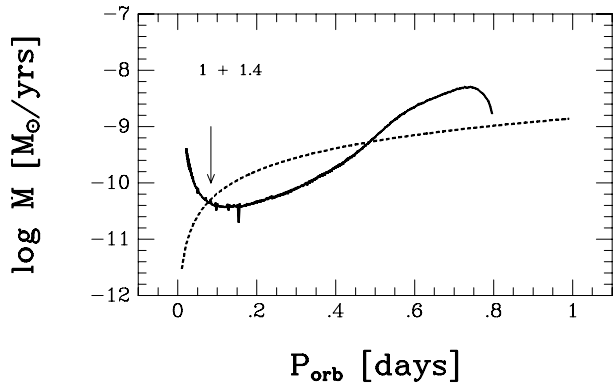,height=5cm,width=8cm}
\end{center}
\caption{Evolution of the mass accretion rate versus orbital period.
 The critical mass accretion rate ( equation (1)) is also shown
 (dashed line). The arrow shows the position of  SAX J1808.4 -- 3658}
\end{figure}

In Fig.1 we have drawn the value of the  critical mass accretion rate
 as a function of $P_\textrm{orb}$. Near an orbital period of two hours the
 calculated mass accretion rate is below the  critical value and hence
 the system can not be a persistent X-ray source.

Chakrabarty \& Morgan (1998) proposed that if the X-ray bright state lasts long enough (or if the pulsations remain detectable in quiescence), further X-ray longer-term timing observations may yield a detection of orbital period evolution, which would
 probe effects of gravitational radiation, tital interactions and mass
 loss.   If the orbital evolution is driven by  gravitational wave
 radiation  then the orbital period will decay according to
 (Landau \& Lifshitz, 1971)
\begin{equation}
\dot{P}_\textrm{orb}=-\frac{192\pi}{5}(\frac {2\pi}{P_\textrm{orb}})^{5/3}
 \frac{M_\textrm{s}}{\msun} \frac{M_\textrm{ns}}{\msun} (\frac{M_\textrm{T}}{\msun})^{-1/3}
(\frac{ G\msun}{c^3})^{5/3}
\end{equation}
where $\frac{G \msun}{c^3}$ = 4.925 $\times10^{-6}$s
 $M_\textrm{s}$ is the mass of secondary and $M_\textrm{T}$= $M_\textrm{s}$
+$M_\textrm{ns}$ is the
 total mass . If the secondary has a mass of 0.05$\msun$ and
 $M_\textrm{ns}$=1.4$\msun$ (as discussed by Chakrabarty \& Morgan, 1998)
 then $\dot{P}_\textrm{orb}$$\sim$$ -$ 8.1 $\times 10^{-14}$.
In our model, since the orbital evolution is driven by magnetic
 braking the orbital period decay is faster ($\sim$$ - $ 4.5 $\times
 10^{-13}$). So the determination of the value of orbital period decay
 may give  important clue to understanding the prehistory of this
 system.
\subsection{{\bf The chemical composition of the secondary and
 the X-ray bursts}}

  In our scenario the secondary is a helium rich star. There are two
 systems, namely MXB 1916 -- 05 ($P_\textrm{orb}$ = 50 min) and MXB 1820--30 ($
P_\textrm{orb}$=11.4 min), which according to
 current understanding of the low-mass binary evolution  with compact
 object as an accretor,  must have a Roche - lobe filling helium rich
 secondary in order  to explain its short orbital period
 (Paczynski \& Sienkewicz ,1981). 

  Computations of  X-ray bursts  at the surface of 
 accreting neutron stars have shown (Ayasly\&Joss,(AJ) 1982,
 Kudrjashov\&Ergma,1983) that a short total duration $\tau_\textrm{d}$  of the burst is
 possible only if the matter  is helium rich (X$\leq$0.2-0.3, where
 X is the hydrogen abundance by mass) before unstable burning starts.
 It may be possible  that during the accretion phase the hot CNO hydrogen
 burning takes place in the accreted matter on a timescale 
$\sim$ $10^2\times$ ($N_\textrm{p}/N_\textrm{CNO}$) s $\approx 10^5$ s. If the recurence
 time between two succesive burst is longer that this hot CNO hydrogen
 burning time the hydrogen may be exhausted  at the bottom of the accreted
 envelope . Crude estimate using the bursts light curve published by in't Zand
et al (1998) gives $\tau_\textrm{d}$ $\sim$ 14 - 15 s.
 Let us analyze the observed data for the X-ray burst using the
 AJ numerical models. Since the accretion rate (from observations) is a
 few times $10^{16}$g s$^{-1}$ then we may compare it with the AJ model
 1 which has been computed with a similar mass accretion rate (Z = 0.01 where 
Z is the metallicity of the accreted matter).
 In this model the hydrogen abundance, $X_\textrm{b}$, at the base of the freshly
 accreted matter prior to the flash is 0.59. The burst total duration ,
 $\tau_\textrm{d}$,
 is 34 s and the burst recurence time, $t_\textrm{rec}$ $\sim$ 4.7 hours .
 The maximum burst luminosity  $L_\textrm{max}$ is  5$\times 10^{37}$erg s$^{-1}$.
 The AJ model 3, in which the mass accretion rate is one order magnitude
 less, has $X_\textrm{b}$  = 0.0004, $t_\textrm{rec} \sim$ 16.3 hours, 
$\tau_\textrm{d}\sim$ 7 s and
 $L_\textrm{max}$ = 2.5$\times 10^{37}$erg s$^{-1}$. The AJ model 16 with the
 same accretion rate as for model 1, but with Z=0.001, has $X_\textrm{b}$ = 0.65,
  $t_\textrm{rec}$ $\sim$ 11.2 hours, $\tau_\textrm{d}\sim$ 38 s and $L_\textrm{max}$ = 7.5$\times
 10^{37}$erg s$^{-1}$.  From these results it can clearly be seen that in
 order to  have a short total duration of the burst  , the hydrogen must either be
 exhausted or the accreted matter must be helium rich in the envelope.
 In SAX J1808.4 -- 3658 the time between two observed successive bursts
 is  $\sim$ 14 hours, which is much longer than for model 1 with a similar
 mass accretion rate. One possible explanation is that the amount of CNO
 elements is reduced due to spallation, as it was discussed by
 Bildsten et al. (1992), and hence the recurrence time will increase
 ( compare AJ models 1 and 16). It is also possible to speculate that the burst
recurence time may increase with increase of the radius of the neutron star
(referee comment). For example in AJ model 14 ($R_\textrm{ns}$=13 km) the recurrence
time is  $\sim$ 16 hours. In our analyze we used AJ standart model results (
$R_\textrm{ns}$ = 6.57 km ) since according to in't Zand et al (1998) a burst
 emitting sphere radius is of 8 km ( at the distance of 4 kpc). Although the neutron
star radius was determined by in`t Zand et al using a simple black body
fit to the burst spectrum and can therefore underestimate the radius of the 
emitting sphere (Titarchuk, 1994).

So to explain  the short total
duration of the burst and the Eddington luminosity we propose that the accreted matter
 must be helium rich . In order to explain long recurrence time one would need either
reduced Z value or large $R_\textrm{ns}$.
  
\subsection{{\bf The possibility of detection of the millisecond
 radiopulsar.}}

Since the millisecond X-ray pulsar SAX J1808.4-3658 has orbital
 parameters very close to those of the eclipsing binary pulsar system
 PSR J2051-0827, it has been proposed 
 (Chakrabarty \& Morgan, 1998, Wijnands \& van der Klis, 1998) that 
 this system may emerge as a radio pulsar after the X-ray  state ends. 
 Lets us estimate the optical depth for the free - free process
 (Illarionov\&Sunyaev,1975)
\begin{equation}
\tau_\textrm{ff}=\frac{100 (\dot{M}/\dot{M_\textrm{Edd}})^2}{M_\textrm{s}
+M_\textrm{ns}}
(\frac{\lambda}{75 cm})^2(\frac{T}{10^4})^{-1.5}(\frac{P_\textrm{orb}}{1yr})^{-2}
\end{equation}
where $\dot{M_\textrm{Edd}}$ is the Eddington mass accretion rate.
 Accepting $\dot{M}$$\sim$ $10^{-11}$$\msun$$yr^{-1}$, T=6000 
(less than the hydrogen ionization temperature),
$M_\textrm{ns}$=1.4 $\msun$,$M_s$ = 0.1$\msun$, $\lambda$= 3 cm and
 $P_\textrm{orb}$= 2 hours we can find that $\tau_\textrm{ff}$ is  $\sim$ 1,
 which means that at $\lambda$ $<$ 3 cm it may be possible to observe
 radio emission. For longer wavelenghts (20 or 70 cm) $\tau_\textrm{ff}$ is
 greater that unity.
\section{Conclusion}
 We propose that the progenitor of SAX J1808.4 -- 3658 was a low mass
 binary with a neutron star and a secondary that filled its Roche- lobe
 at the turn-off Main-Sequence  .During the bright X-ray phase the neutron
 star will spin-up to a millisecond period. Due to a chemical composition
 gradient, the secondary does not become fully convective when its mass
 has decreased below 0.3 $\msun$ and magnetic braking is therefore still
 the most effective mechanism to remove orbital angular momentum.  Near an
 orbital period of two hours  the mass transfer rate is so small
 ($\sim$ $10^{-11}\msun$y$r^{-1}$) that the system can not be a
 persistent X-ray source. In our evolutionary scenario, due to deep mixing,
 the secondary is a helium rich star. Since the accreted matter is helium
 rich it is easy to expain the observed short total duration
of the burst .
 It is estimated that in this evolutionary picture it may be possible to
 observe radio emission from the pulsar only at short wave lenghts,
 $\lambda$$<$ 3cm. Our model predicts a faster orbital period decay than
 in the case where  the orbital evolution is  driven only by the
 gravitational  radiation.      
\begin{acknowledgements}
EE would like to thank  Profs. Jan van Paradijs and Ed van den Heuvel,
 Drs. Stapper and Wijands for comments and suggestions that helped us
 improve this paper. We should like to thank anonymous referee for very
useful comments on the earlier version of the paper.
 EE ackowledges support through NWO Spinoza grant
 to E.P.J. van den Heuvel. EE acknowledges the warm hospitality of the
 Astronomical Institute ``Anton Pannekoek'' where this study was
 accomplished.  This work has been supported by the Estonian Science
 Foundation grant N 2446. 
\end{acknowledgements}
{}
\end{document}